# Angular emission of scintillators for nuclear fusion diagnostics

M. Rodríguez-Ramos[a, *], J. García-López[a, b], M. Videla-Trevin[a, ‡], J. González-Martin[c], P. Alvarez-Frau[b], M. Kocan[d]

[a]Centro Nacional de Aceleradores (U. Sevilla, CSIC, J. de Andalucia).
[b]Dpto. de Física Atómica, Molecular y Nuclear, Universidad de Sevilla, 41012, Seville, Spain.
[c]Department of Mechanical and Manufacturing Engineering, University of Seville, Seville, Spain.
[d]ITER Organization, Route de Vinon sur Verdon, CS 90 046, 13067 Saint-Paul-lez-Durance, France.
[‡]Currently not affiliated.
[*]Corresponding author: mrodriguez67@us.es

**Keywords**: Scintillators, Nuclear fusion reactor, Fast Ion Loss Detectors, Angular ionoluminescence emission.

## ABSTRACT:

Accurate characterization of fast-ion behavior is essential for the safe and efficient operation of nuclear fusion plasmas, as energetic particle losses can degrade plasma performance and damage reactor components. Scintillator-based detectors are widely employed to monitor fast ions; however, existing studies often assume isotropic light emission, neglecting potential angular dependencies that can compromise the determination of ion fluxes. In this work, we investigate the angular emission properties of two commercial scintillators, TG-Green and β-SiAlON, under irradiation with 3.5 MeV $He^{++}$ and 1 MeV $D^+$ beams, representative of conditions in future fusion devices such as ITER. A novel experimental setup, combining precise optical alignment, angular scanning, and rigorous calibration, was developed to measure the detection efficiency as a function of observation angle. Prior to the characterization, stability tests demonstrated negligible radiation-induced degradation under the applied fluences, and transmission losses due to optical fiber bending were found to be below 1.5%. The results reveal a pronounced angular anisotropy in scintillation emission for both materials, with intensity decreasing as the detection angle increases, well described by an empirical cosine-based model. Additionally, the normalized response shows minimal dependence on ion species or energy. These findings improve scintillator-based diagnostics, allowing more accurate measurement of fast-ion fluxes in fusion plasmas.

## 1. Introduction:

In nuclear fusion plasmas, the loss of energetic particles, referred to as fast ions, poses a significant challenge by potentially damaging the structural integrity of the device and reducing plasma performance [1,2]. To mitigate this issue, advanced diagnostic tools have been developed to monitor fast ion behavior, with scintillator-based detectors being a key technology [3,4]. These detectors, widely employed in current tokamaks, rely on scintillator materials that emit light when struck by energetic particles. The scintillation process enables the detection of ion impacts, with the ion's trajectories and their strikes on the detector plate being dependent on their gyroradius and pitch angle, thus providing valuable insights into their local velocity-space distribution [5,6]. A comprehensive understanding of scintillator materials under relevant conditions, such as fusion-relevant ion species, energy levels in the MeV range, fluences, and temperature, remains critical for optimizing diagnostic systems.

Although several studies have characterized scintillators under these conditions [7], most have assumed isotropic light emission, neglecting potential angular dependencies. This assumption can lead to a misestimation of the total light emitted by the scintillators. Understanding the dependence of light emission on the detection angle is essential for accurately modeling and determining the efficiency of scintillator materials, which is a key parameter in quantifying the absolute flux of fast ions escaping from the plasma in a fusion reactor [2,8]. This study investigates the angular emission characteristics of two commercial scintillator materials, TG-Green and β-SiALON, employed in nuclear fusion diagnostics [4,9], thereby bridging a critical gap in the understanding of their luminescent properties. The luminescence of these materials was studied using a 3.5 MeV He and 1 MeV D beams, ionic species and energies relevant for future nuclear fusion reactors like ITER [10]. The relative emission yield as a function of detection angle was measured for each material at room temperature. To facilitate these measurements, a new experimental setup based on an optical system and measurement protocol was developed at the multipurpose chamber of the National Center for Accelerators (CNA), allowing for a more precise characterization of scintillator behavior. This paper is organized as follows: the introduction outlines the importance of scintillator materials in nuclear fusion diagnostics and identifies a gap in the literature regarding their angular emission properties. Section 2 describes the experimental setup used for ionoluminescence measurements and determining the relative emission yield, in terms of counts per nA as a function of detection angle. Section 3 discusses the calibration process of the spectrometer in terms of wavelength, which is crucial for ensuring accurate measurements. Section 4 presents different mechanisms that can affect the light measured by our acquisition system, such as optical fiber bending and/or scintillator degradation caused by ions, which must be distinguished from any variations associated with the angular emission of the scintillators. Section 5 presents the main results, including the response of the scintillators at different angles, the determination of the relative emission yield using a cross-correlation algorithm, and a polar histogram plot of the normalized intensity for the two scintillator samples. Finally, the conclusions summarize the key findings and their implications for future diagnostics in nuclear fusion reactors.

2. **Experimental setup:**

In this study, two different scintillators were selected based on their common use in nuclear fusion diagnostics, such as TG-Green [11], or for their excellent luminescent properties at high temperatures, as in the case of β-SiAlON [12].

- TG-Green: This phosphor, manufactured by Sarnoff Corporation (USA), is a mixed $SrGa_2S_4$ host doped with $Eu^{2+}$ and has a density of 3.65 g/cm$^3$. It is well known for its green emission peak at ~536 nm, arising from the allowed $4f^65d^1 \rightarrow 4f^7$ electronic transition of $Eu^{2+}$ centers. TG-Green exhibits a bandgap of 4.2 eV at 300 K and a fast response time of ~490 ns [13].

- β-SiAlON: Derived from silicon nitride ($Si_3N_4$) by partial substitution of Si and N with Al and O, this ceramic material acquires phosphorescent properties when doped with $Eu^{2+}$. Its emission typically lies in the yellow-green range (450-550 nm), with a main peak at ~492 nm due to the allowed $4f^65d^1 \rightarrow 4f^7$ transition of $Eu^{2+}$ ions [14].

Known for its high mechanical strength and thermal stability, β-SiAlON can withstand temperatures above 1000ºC and is resistant to significant wear and corrosion [15].

The TG-Green was commercially coated on a stainless steel substrate with dimensions of 2.5 cm x 1.5 cm, while the β-SiAlON sample has dimensions of 2.2 cm x 2.2 cm and was deposited by precipitation at the Material Science Laboratory at the engineering school of the University of Seville [16]. For the last one, the deposition was carried out by preparing a colloidal solution through dispersing the scintillator powder in distilled water, with the addition of a small amount of deflocculant. The pH of the suspension is adjusted to achieve an optimal z-potential for more uniform layers. The suspension is pipetted onto the stainless-steel plate and left to dry overnight in an incubator set at a controlled temperature of 50ºC. The characterization of the samples was performed at the multipurpose beamline of the CNA using a 3 MV tandem accelerator, capable of delivering a wide range of ion species and energies relevant to nuclear fusion applications [17]. All experiments were conducted under high-vacuum ∼1x $10^{-6}$ mbar) conditions to minimize beam scattering and surface contamination during the irradiation. Both scintillators were irradiated with ion beams of 3.5 MeV $He^{++}$ and 1 MeV $D^+$. The selected energy corresponds to the alpha particles produced in the D-T fusion reaction [18], whereas the latter represents the characteristic energy of the neutral beam injector that will be implemented at ITER for auxiliary plasma heating [19]. The scintillator samples were mounted on an electrically isolated holder coupled to a precision current integrator (Model 439 from ORTEC) to continuously monitor the beam current incident on the samples, which is required for normalization in order to accurately determine the scintillator efficiency. The holder was equipped with computer-controlled stepping motors that allowed independent horizontal and vertical displacements with an accuracy of 0.1 mm, thereby enabling fine adjustment of the irradiation spot. To suppress the emission of secondary electrons generated during the irradiations, which could otherwise lead to systematic errors in the beam current measurements [20], a positive bias voltage of +200 V was applied to the sample holder throughout the experiments. The system also permitted rotation of the samples to guarantee that the ion beam impinged at normal incidence. Fig.1 presents a photograph of the sample holder with the two scintillator samples used in this study. To facilitate the alignment of the ion beam and optimize its parameters, a quartz ($SiO_2$) piece was also mounted on the holder. The beam spot size was set to 3 mm in diameter to reduced the current density to avoid the degradatation of the targets using a set of collimators positioned at the entrance of the vacuum chamber.

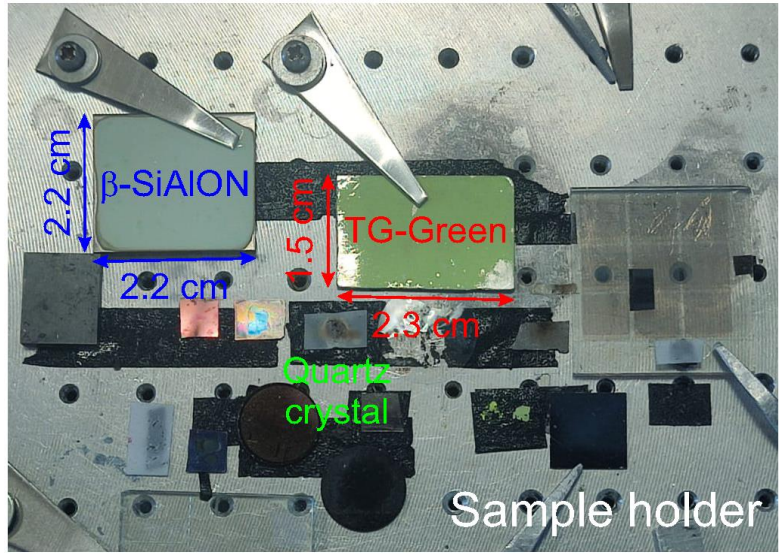

Figure 1: Scintillator samples (TG-Green and β-SiAlON) mounted on the target holder before angular measurements, with a SiO$_2$ crystal used to set the ion beam parameters and determine its position.

Ion beam currents were maintained between 1-10 nA, sufficient to enhance light emission while avoiding degradation of the scintillator's spectral response due to radiation damage. Because the ionoluminescence signal depends on the beam current, it was crucial to keep the current approximately constant throughout the measurements, although oscillations in the current profile, an effect inherent to the ion source, are particularly relevant for He$^{++}$ beams. All irradiations were performed at room temperature and for accuracy results under dark ambient conditions. The optical acquisition system consisted of two optical fibers and an optical spectrometer. One end of the first optical fiber (#1, 1 mm diameter from Hangzhou ZS) was installed inside the vacuum chamber at the platform used to hold the passivated implanted planar silicon detector to applied the Rutherford backscatering technique. This holder allowed the fiber head to be positioned at various angles, using a rotary manipulator capable of rotation from 0º (this direction corresponds to ion beam axis) to 90º respect to the ion beam axis with 0.5º precision. To secure the fiber, it was placed in a manipulator within a plastic casing manufactured by 3D printing. The other end of fiber #1 was connected to a dedicated vacuum feedthrough. An SMA connector on the external side of the feedthrough allowed connection to a second optical fiber (#2, 1 mm diameter from OceanOptics), which was then connected to the input of the optical spectrometer (Ocean Optics QE65000 Pro). The spectrometer was equipped with a 300 grooves/mm grating, covering a spectral range from 200 to 1000 nm and providing a spectral resolution of 1-2 nm. To ensure accurate alignment of the optical fiber with the ion beam impact point on the samples, a laser was coupled to the external end of fiber #2. This setup allowed fine adjustments of the fiber position, ensuring reliable alignment before measurements. A schematic of the experimental layout and main components is presented in Fig.2a. Fig.3a shows a photo of the optical fiber #1 positioned inside the 3D-printed housing.

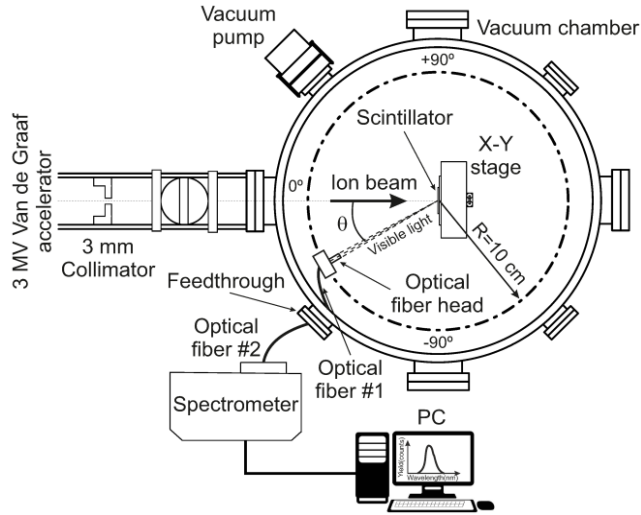

Figure 2: Schematic of the experimental setup used to measure the angular emission of scintillators under ion beam irradiation. The inner optical fiber (#1) is mounted to rotate around the ion beam axis.

Although the rotary system allows full range of angular rotation, the dimensions of the plastic housing (4x4x2.7 cm$^3$) containing the optical fiber only permit measurements from 30º up to 60º. For smaller angles, the ion beam is blocked by the plastic casing, preventing accurate measurements by affecting the current reaching the scintillator. The effect of the ion beam impact on the housing that holds the optical fiber, visible as a darkened region caused by ion-induced damage during one of the experimental tests, can be observed in Fig.3b.

3. **<u>Analytical proccedure:</u>**

Prior to the characterization of the angular emission of the samples, the following steps need to be considered:

- 3.1 Wavelength calibration

To improve the quality of the experimental results, the first step must be the wavelength calibration of the spectrometer, since the wavelengths in commercial spectrometers can drift slightly over time and under varying environmental conditions [21]. Without proper calibration, the spectral data obtained from the instrument may exhibit systematic errors, leading to inaccurate identification of emission or absorption lines and compromising the validity of subsequent analyses. In this work, the wavelength calibration of our optical acquisition system was performed using a pencil-style low-pressure mercury-argon (Hg(Ar)) discharge lamp (model 6035 from Oriel [22]), which produces first-order mercury and argon atomic emission lines covering the visible spectral range. The procedure consists of placing the Hg(Ar) lamp near an optical fiber connected to the spectrometer and recording the emission spectrum. Before the measurement, the lamp should be allowed to warm up for five minutes to ensure accurate readings. The acquisition was carried out using the SpectraSuite software setting the exposure time to 8 ms, which places the most intense peaks at the spectrometer's saturation limit (16-bit dynamic range, corresponding to a theoretical maximum of 65536 counts) while still providing an adequate signal-to-noise ratio to resolve the weaker peaks required for calibration. To increase the stability of the recorded spectrum, an average of 100 spectra was used.

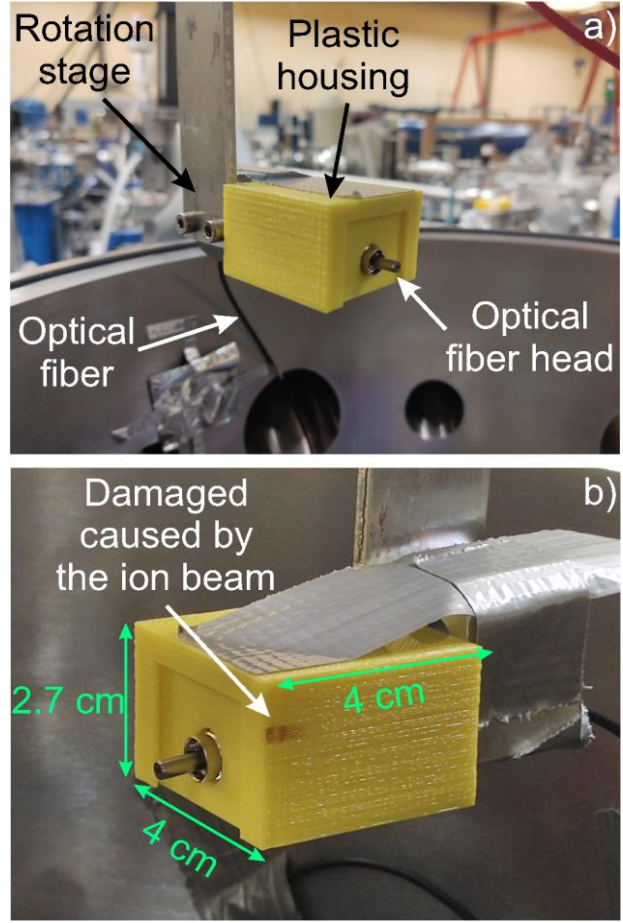

Figure 3: a) Plastic casing for the inner optical fiber installed on the rotation stage in the multipurpose chamber at CNA. b) The casing showing a noticeable burn mark caused by ion beam impact at lower angles. The dimensions of the piece (4x4x2.7 cm$^3$) restrict the angular scan from 30° to 60°.

The averaged spectrum of the calibration lamp is shown in Fig.4(Top). The main spectral lines are clearly visible, ranging from 310 nm to 850 nm, with the most intense line at ~435 nm in our spectrometer. These lines are slightly shifted relative to the tabulated wavelengths of the Hg(Ar) lamp provided by the supplier, indicating that the spectrometer requires calibration. A zoomed-in view of the region close to the infrared containing the weaker peaks is included as an inset. Following the spectrometer protocol for the wavelength calibration [23], the relevant parameter is the wavelength position of the spectral lines rather than their intensity. To establish the relationship between pixel position and wavelength, the centroids of the principal Hg(Ar) emission lines were accurately determined within the recorded spectrum. These centroid positions, denoted as pixel coordinates $p_i$, were then mapped to their corresponding standard wavelengths $\lambda_i$ from the known lamp emission lines. This produced a discrete dataset of calibration points $p_i$, $\lambda_i$). Due to the inherent optical properties and dispersion characteristics of the spectrometer's grating and lens system, the pixel-to-wavelength relationship is generally non-linear. In a diffraction-grating spectrometer, the fundamental relation between the wavelength $\lambda$ and the diffraction angle $\theta_m$ is given by the grating equation [24,25]:

$$m\lambda = d(\sin(\theta_i) + \sin(\theta_m))$$

where m is the diffraction order, d is the grating spacing, and $\theta_i$ and $\theta_m$ are the angles of incidence and diffraction, respectively. The position x of the diffracted light on the detector is related to the diffraction angle through the optical system, which for a lens of focal length f can be expressed as:

$$x = f \tan \theta_m$$

Substituting $\theta_m$ from the grating equation, we obtain:

$$x(\lambda) = f \tan\left[\arcsin\left(\frac{m\lambda}{d} - \sin\theta_i\right)\right]$$

For small diffraction angles and moderate spectral ranges, it is convenient to expand tan(arcsin y) in a Taylor series:

$$x(\lambda) \sim f\left[\left(\frac{m\lambda}{d} - \sin\theta_i\right) + \frac{1}{6}\left(\frac{m\lambda}{d} - \sin\theta_i\right)^3 + \cdots\right]$$

By inverting this relation to express the wavelength as a function of detector position x, and converting to pixel number p via $p \sim x/\Delta x$, a natural functional form emerges. This indicates that the wavelength dependence on detector position can be effectively approximated by a polynomial expansion. In practice, additional optical imperfections (e.g., aberrations or alignment errors) introduce even-order contributions, so the pixel-to-wavelength calibration is typically performed using a polynomial of degree two to four. For our optical spectrometer, the relationship between pixel number and wavelength was modeled using a third-order polynomial, which can be expressed as:

$$\lambda_p = \sum_{k=0}^{3} C_k p^k = C_0 + C_1 p^1 + C_2 p^2 + C_3 p^3$$

where $\lambda_p$ is the wavelength corresponding to pixel p, $C_0$ accounts for the wavelength corresponding to the first pixel (offset), $C_1$ represents the primary linear dispersion, $C_2$ and $C_3$ compensate for optical aberrations, misalignments, and non-ideal grating behavior. The wavelengths of the main emission lines as a function of their pixel position can be seen in Fig.4(Bottom). It can be observed that higher wavelengths correspond to higher pixel numbers. All polynomial coefficients were determined via a least-squares regression fit of the calibration dataset and were subsequently entered into the spectrometer configuration. From Fig.4(Bottom), the fitted calibration curve shows excellent agreement with the known emission lines. These results confirm that the calibration procedure effectively corrects systematic deviations and nonlinearities in the spectral response. To investigate the angular dependence of the scintillation light emission, it is essential to account for all physical processes that may influence the intensity measured by the detection system. Both intrinsic and extrinsic factors can modify the detected optical signal. Intrinsic effects are related to the scintillator material itself, such as radiation-induced damage that reduces the light yield by generating color centers and structural defects capable of absorbing or scattering the emitted photons [26]. Extrinsic factors, on the other hand, arise from the optical collection and transmission system.

In optical fibers, light attenuation can result from several mechanisms, including intrinsic material absorption, Rayleigh scattering due to microscopic inhomogeneities in the fiber core, and additional losses associated with impurities, defects, or imperfect coupling between components. Mechanical stresses such as bending or microbending can also lead to localized scattering and mode losses [27], while variations in temperature or tension may alter the refractive index and affect light propagation stability [28]. To ensure that any observed variations in emission intensity truly reflect the intrinsic angular dependence of the scintillation process, it is necessary to first quantify and minimize the influence of these extrinsic and intrinsic sources of variability. In this work, particular attention is given to two critical factors: macrobending losses in the optical fiber and the degradation of ionoluminescent emission caused by radiation-induced damage. Both mechanisms can significantly alter the detected light intensity and must therefore be carefully characterized to enable an accurate interpretation of the angular emission behavior.

- 3.2 Effect of bending the optical fiber on signal transmission

For the experimental setup considered in this work, the most critical factor to assess is macrobending, since the inner fiber must rotate around the ion beam axis during the angular measurements. Large bending radii can cause significant light leakage from the core, reducing the effective signal and potentially compromising the operational stability of the collected data. To evaluate the effect of fiber bending, a preliminary test was performed using a light source rigidly fixed to the optical fiber. The optical fiber was mounted on the rotation holder, and a light-emitting diode (LED, supplied by Shishixiaodian, with a main emission centered at ∼517 nm) was positioned in front of the fiber such that it rotated together with the fiber as the rotation holder was turned. This configuration ensures that the relative geometry between the fiber and the LED remains constant for all rotation angles, so that any variation in the measured intensity can be attributed solely to fiber bending.

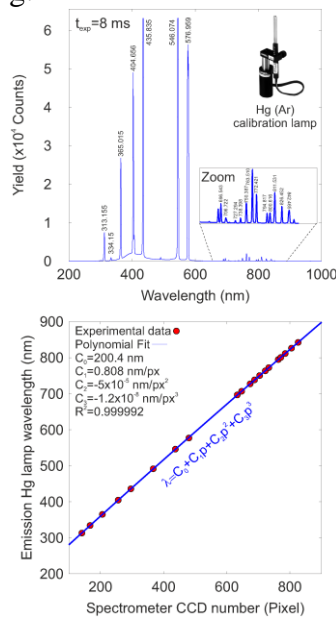

Figure 4: a) Averaged emission spectrum of a Hg(Ar) lamp, with the main atomic lines clearly visible. The numbers above each peak indicate the central wavelengths provided by the manufacturer. Inset shows a photograph of the Hg(Ar) lamp. b) Wavelengths of the principal emission lines plotted against CCD pixel positions (red dots), with the blue line showing the best third-order polynomial fit.

The LED was powered by a 3.3 V button cell (Model CR2032 from Panasonic), and the emitted light was recorded using the spectrometer with an exposure time of 20 ms to avoid saturation of the main emission. Prior to this test, the long-term stability of the LED emission was evaluated over a 5-minute period. Fig.5a shows the spectra series recorded during this time interval with the spectrometer set to 100 ms. A progressive decrease in signal intensity is clearly observed, evidenced by the reduction in the spectral peak height. This effect becomes even more evident when representing the measured light intensity as the integrated area of the emission spectrum within the region of interest from 460 to 640 nm, plotted as a function of time. Fig.5b shows the normalized intensity of the green LED over the measurement period. A continuous decrease in emission is observed, with the brightness dropping by approximately 5% relative to the initial emission after 300 s. This drop in LED intensity is attributed to the button cell used to power the source: as the battery voltage gradually decreases, it delivers a lower current, which in turn reduces the LED brightness. It is important to note that, although the LED current decreases over time, the intensity decay follows a nearly constant rate, allowing for a post-processing correction to compensate for the emission loss. A linear fit to the experimental data shows a reduction rate of approximately ~0.017 % s. To evaluate the effect of fiber bending, emission spectra from the LED were recorded during a continuous scan ranging from 0° to 60° in the clockwise direction, followed by a return scan from 60° back to 0° (counterclockwise rotation). The total time required to complete both angular sweeps was less than 80 s, ensuring that any temporal drift in the LED intensity can be corrected using the previously determined reduction rate of the LED emission. An offline data analysis was performed using MATLAB to extract the angular intensity profile. First, the background level of each spectrum was subtracted by taking the mean intensity within the wavelength range [240, 340] nm, a region where no LED emission is expected. The LED intensity was then quantified as the integrated spectral area within [450, 610] nm, corresponding to the main emission band of the light source. The resulting angular intensity profiles for both the forward (0º → 60º) and reverse (60º → 0º) scans are presented in Fig.5c. From these results, the following conclusions can be drawn. During the clockwise rotation (blue line), a decrease in the measured intensity is observed as the angular position of the fiber increases up to 60°, corresponding to a reduction of 2.05% with respect to the initial position. At 60°, an abrupt difference of approximately 2.39% appears between the intensity measured at the end of the clockwise scan and the value obtained at the beginning of the counterclockwise scan. This discontinuity is not related to additional bending losses but is instead explained by the experimental procedure and by the intrinsic attenuation of the LED brightness over time. After completing the clockwise scan, the acquisition was stopped, and approximately 140 s were required to reconfigure the spectrometer for the counterclockwise measurement. During this interval, the LED brightness decreased by about 140x0.017~2.38%, which matches the observed drop in intensity almost exactly, confirming that the effect is solely due to LED emission drift and not to the bending of the optical fiber. During the counterclockwise rotation (red line), the measured intensity gradually increases as the fiber approaches 0°, resulting in a 0.8% enhancement relative to the initial position. Taking into account the time required to complete each scan, the measured signal was corrected to compensate for the gradual decrease in LED brightness. Fig.6a shows a polar plot of the normalized and corrected intensity (referenced to the first point at 0º), covering the range from 0º to 60º in both clockwise and counterclockwise directions, with angular steps of 5º. From the polar diagram, a slight reduction in the recorded intensity is observed as the angle increases.

When the fiber is subsequently rotated counterclockwise, the measured intensity returns close to its initial value at 0°, confirming the reproducibility of the measurement during the reverse rotation. Assuming that the intensity variations caused by the decreasing LED brightness and those associated with bending losses are independent and multiplicative [29], the signal measured by the spectrometer can be expressed as:

$$ESP(t) = \frac{I_{LED}(t) \times B(\theta(t))}{100}$$

where $ESP(t) \equiv ESP(\theta)$ represents the relative measured intensity (in %) at time t, corresponding to an angular position θ. In our analysis, the normalization criterion $ESP(t=0)=ESP(\theta =0°)=100\%$ was adopted for the initial reference point. The term $I_{LED}(t)$ denotes the relative LED intensity (in %) as a function of time. Long-term stability measurements indicate that the LED emission can be approximated by:

$$I_{LED}(t) \sim 100 - 0.017 \times t$$

with t expressed in seconds. Finally, B(θ) represents the percentage factor quantifying the bending-induced loss at the time associated with angle θ. By definition, B(0°)=100%, since the initial position introduces no bending, and all other values are expressed relative to this reference. From the expression above, the bending factor for an angle θ measured at time t can be expressed as:

$$B(\theta(t)) = 100 \times \frac{ESP(t)}{I_{LED}(t)}$$

Fig.6b illustrates the bending factor during both clockwise and counterclockwise rotations. The bending response exhibits a non-monotonic trend, and while a slight asymmetry due to fiber deformation is noticeable, the experimental data demonstrate that, for our setup and the optical fiber used, transmission intensity variations caused by bending remain below 1.5% across the angular range considered.

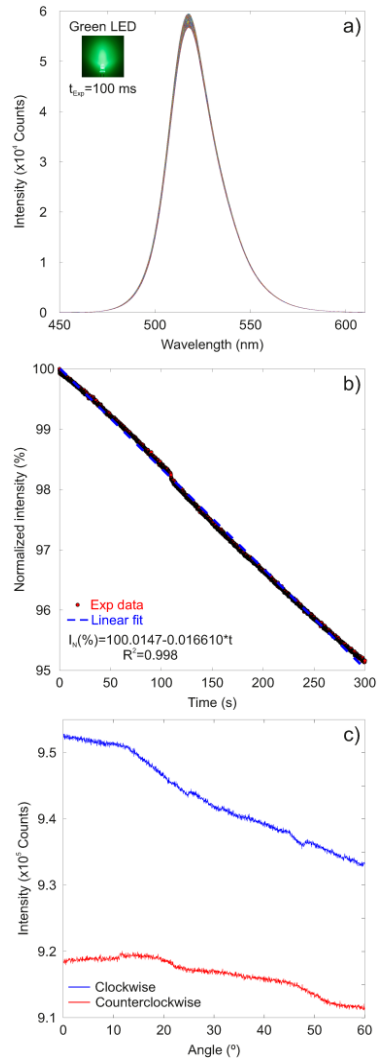

Figure 5: a) Green LED spectra measured over a total duration of 300 s, with the spectrometer exposure time set to 100 ms. b) Temporal evolution of the normalized LED brightness (red points), showing a continuous decrease at a fixed rate. The blue line represents the best linear fit. c) LED intensity as a function of rotation angle during the clockwise (blue) and counterclockwise (red) angular scans. The drop at 60° between scans is due to the gradual decrease in LED brightness during the time interval between measurements.

Consequently, the influence of bending can be considered negligible. This minimal effect indicates that any changes in the emitted light with angular position observed arise primarily from the luminescent properties rather than from transmission losses associated with fiber bending.

- 3.3 Scintillator emission stability.

Finally, to ensure that any observed decrease in emitted light intensity could be attributed to angular emission dependence rather than to radiation-induced quenching effects, both scintillators were subjected to a pre-irradiation phase. This procedure aimed to evaluate the emission behavior as a function of accumulated fluence prior to the angular dependence measurements. During testing, the optical fiber was positioned at an angle of 30º relative to the ion beam axis. The beam was collimated to a diameter of 3 mm and maintained at a current of 4-6 nA for the 3.5 MeV $He^{++}$ irradiation.

Using the central current value of 5.0 nA and assuming a circular beam cross section (diameter 3 mm), this corresponds to a current density of approximately 70.7 nA/cm$^2$ for at least 150 s. For the 1 MeV D$^+$ irradiations, beam currents in the range of 8-11 nA were employed; using the central current value of 9.5 nA yields a current density of approximately 13.4 nA/cm$^2$ for at least 100 s. The emission spectra were recorded using the spectrometer with an exposure time of 1 s per acquisition.

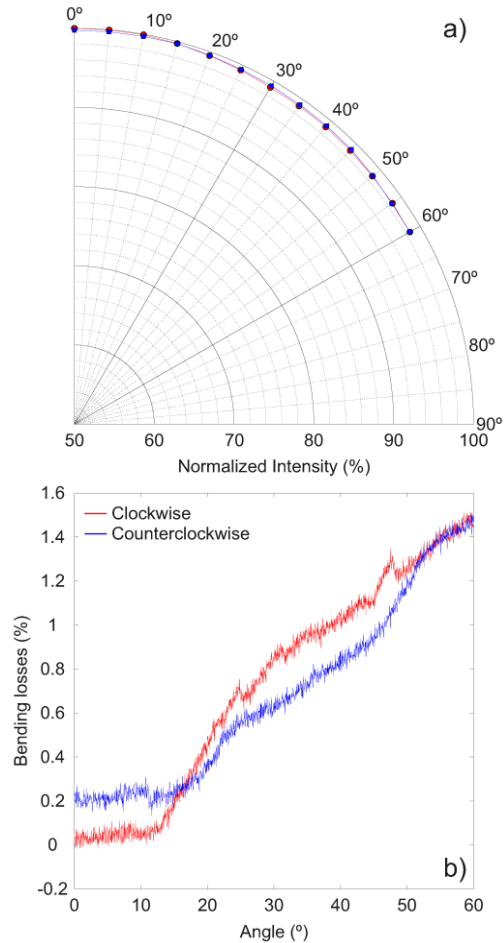

Figure 6: a) Polar plot of the normalized green LED intensity from 0° to 60° for clockwise (red) and counterclockwise (blue) rotations. b) Fiber bending losses (%) as a function of rotation angle for both scans, with maximum losses remaining below 1.5%.

Over the total fluence accumulated during the angular scans, no significant degradation of the scintillation light output was detected, indicating that the radiation dose was insufficient to cause notable damage in either material. The background level of each spectrum was subtracted by averaging the signal intensity within the 240-340 nm wavelength interval, where no emission is expected according to the spectral response of the scintillators. The scintillation intensity was then quantified as the integrated spectral area within 450-540 nm for the β-SiAlON sample and 470-630 nm for TG-Green, corresponding to the principal emission bands of each scintillator, respectively. Fig.7a presents the normalized emission spectra of the two scintillators characterized in this study.

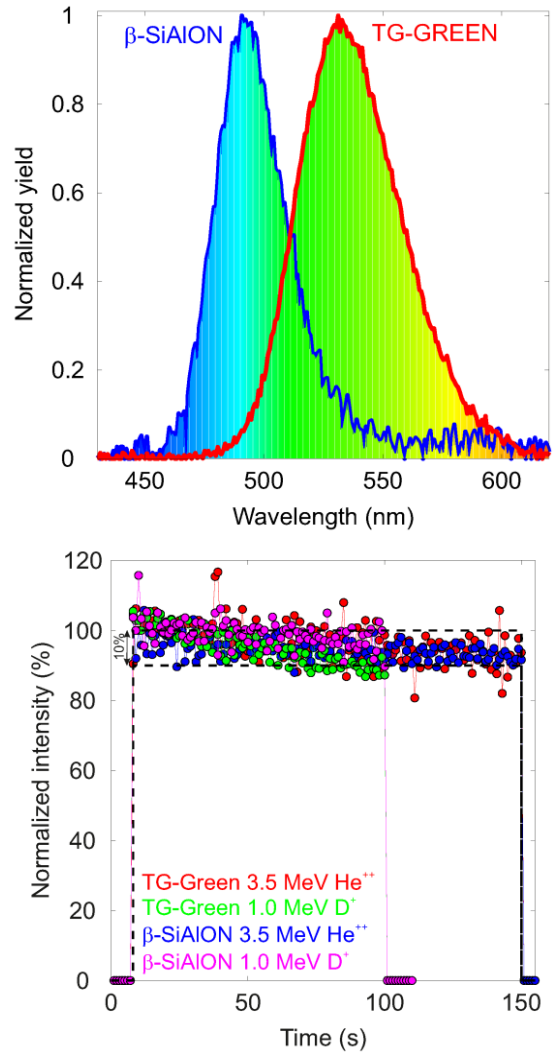

Figure 7: a) Normalized ionoluminescence spectra of TG-Green and β-SiAlON scintillators. b) Temporal evolution of the normalized efficiency under irradiation: TG-Green with 3.5 MeV He$^{++}$ (red), TG-Green with 1 MeV D$^+$ (green), β-SiAlON with 3.5 MeV He$^{++}$ (blue), and β-SiAlON with 1 MeV D$^+$ (magenta).

Both materials exhibit well-defined emission bands within the visible region. The TG-Green scintillator displays a relatively broad emission spectrum, with a maximum centered at approximately 531 nm (2.33 eV), which is consistent with the allowed 4f →5d levels of Eu$^{2+}$. In contrast, β-SiAlON exhibits a narrower emission band with its main peak centered around 492 nm, corresponding to the 4f$^6$5$^d$→4f$^7$ transition of Eu$^{2+}$. Fig.7b shows the evolution of the normalized luminescence intensity, defined as the integrated area of the emission spectra as a function of the irradiation time. It is important to note that the measured intensities were corrected for variations in the ion beam current to account for possible fluctuations during the experiments. From these results, it is evident that the ionoluminescence intensity remained relatively stable throughout the duration of the measurements. Only the TG-Green sample irradiated with deuterium exhibited a moderated decrease in intensity, below 10%. Overall, the results indicate that the luminescent efficiency of both scintillators was largely unaffected by the ion fluence accumulated during the experimental timeframe.

- 3.4 Cross-correlation technique for signal alignment

The ionoluminescence acquisition system at the CNA currently lacks a trigger signal, requiring the manual initiation of both the spectrometer acquisition and the current integrator during the angular measurements. This manual operation introduces a temporal delay between the signals recorded by the two systems, as illustrated in Fig.8a. The figure shows an example of the temporal evolution of the light intensity measured by the spectrometer, together with the ion beam current, during the irradiation of the β-SiAlON sample with a 1 MeV deuterium beam and the optical fiber positioned at θ=45°. Consequently, the accuracy of the measurements depends critically on the precise temporal alignment of these signals, which is necessary to properly normalize the emission intensity. To address this issue, a MATLAB script implementing a cross-correlation algorithm was developed. This algorithm quantifies the similarity between two time-dependent signals as a function of their relative shift, allowing the determination of the optimal temporal delay that maximizes the correlation and achieves the best alignment between them. The discrete cross-correlation function between two finite-length signals x[n] and y[n] is defined as [30]:

$$r_{xy}[k] = \sum_{n=0}^{N-1} x[n]y[n-k]$$

where $r_{xy}[k]$ represents the correlation between x[n] and a lagged version of y[n] by k samples. The variable k denotes the lag (positive or negative), and the summation is taken over the overlapping indices of the two signals. In this work, the cross-correlation was computed using the MATLAB built-in function xcorr without normalization, corresponding to the previous equation. The lag value k at which $r_{xy}[k]$ reaches its maximum indicates the relative time delay between the two recorded signals. For the considered example, Fig.8b shows the cross-correlation function as a function of the temporal delay, indicating that the maximum occurs at an optimal delay of 3 s. This delay was subsequently used to temporally align the data sets before quantifying the emission yield, as shown in Fig.8b. After correcting for the temporal offset between the signals, the emission yield was calculated as the ratio of the luminescence intensity to the ion beam current. The average yield was taken as the mean value of the efficiency profile, and the standard deviation was used to quantify the associated uncertainty, as displayed in Fig.8d. This procedure was applied to obtain the intensity in both scintillators, for the ionic species and energies considered, and for all positions of the angular scan.

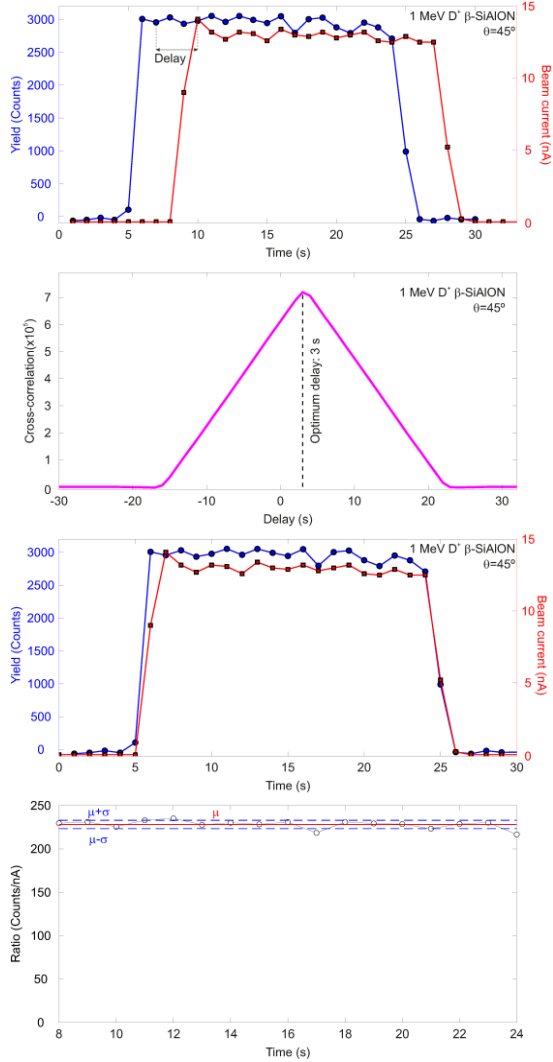

Figure 8: a) Temporal profiles of the ion beam current (red) and the measured intensity (blue) for β-SiAlON irradiated with a 1 MeV D$^+$ beam and with the optical fiber positioned at θ=45° before the alignment correction of both signals. b) Cross-correlation as a function of temporal delay, with the peak indicating the optimal delay for alignment. c) Temporal profiles of the ion beam current (red) and measured intensity (blue) after alignment correction. d) Temporal profile of the efficiency, with the mean value representing the efficiency and the standard deviation indicating the uncertainty.

### 4. Results:

The experimental intensity as a function of the optical fiber's angular position are presented in Fig.9a for the TG-Green and Fig 9b for the β-SiALON for both ionic species and energies. From these graphs a clear angular anisotropy is observed in all scintillator materials tested. Specifically, the measured light intensity decreases gradually as the detection angle increases, indicating that the scintillation response is strongly dependent on the observation direction. This trend is consistently observed across different scintillator materials, ion species, and ion energies, highlighting the robustness and reproducibility of the observed angular dependence under varying experimental conditions.

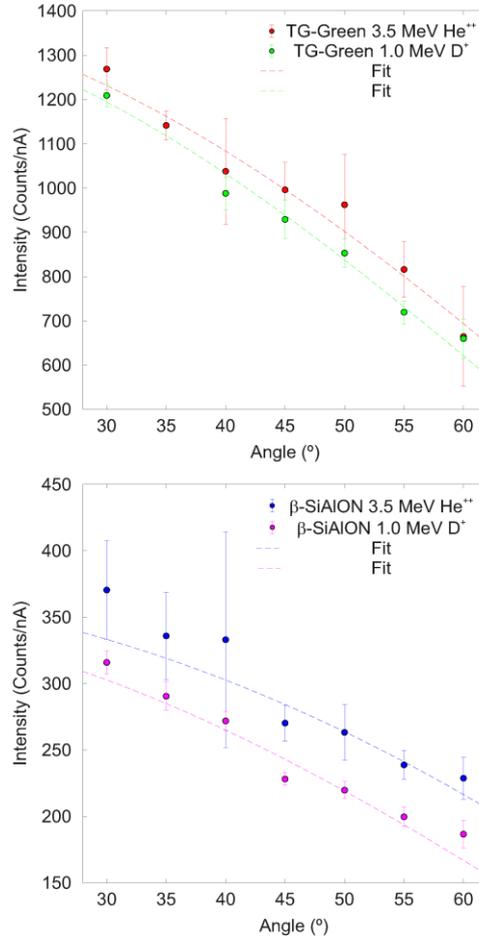

Figure 9: Measured efficiency as a function of detection angle for different ion species and beam energies: TG Green (top) and β-SiAlON (bottom). Dashed lines show the fit of the experimental data to a cosine power-law model.

Comparing different scintillator materials, TG-Green exhibits a higher light yield than the β-SiAlON target. To quantitatively describe the angular dependence, the experimental data were fitted using an empirical model [31]:

$$I(\theta) = A|\cos(\theta)|^n$$

where $I(\theta)$ represents the measured intensity at angle $\theta$, A is a scaling constant, and n is an exponent that captures deviations from a purely cosine dependence. For the fitting procedure, the Levenberg-Marquardt algorithm was used [32,33]. The fit was applied to all measured datasets, yielding generally good agreement with the experimental data, whose results are shown in Table 1.

| Test | A (Counts/nA) | n | $R^2$ |
|---|---|---|---|
| TG-Green - 3.5 MeV $He^{++}$ | 1430 ± 30 | 1.05 ± 0.09 | 0.96 |
| β-SiAlON - 3.5 MeV $He^{++}$ | 373 ± 24 | 0.79 ± 0.13 | 0.84 |
| TG-Green - 1.0 MeV $D^+$ | 1420 ± 40 | 1.19 ± 0.07 | 0.98 |
| β-SiAlON - 1.0 MeV $D^+$ | 354 ± 20 | 1.09 ± 0.16 | 0.94 |

Table 1: Fitting parameters obtained from the angular dependence measurements for different scintillators and ion beam conditions.

The exponent n was consistently found to be close to unity within the limits of experimental uncertainty, suggesting that a simple cosine law provides an adequate approximation for practical purposes. Minor deviations from the expected cosine behavior, particularly for the β-SiAlON scintillator irradiated with a 3.5 MeV helium beam, can be attributed to fluctuations in the ion source, especially during the measurement at 40º, which complicated the cross-correlation corrections required for precise intensity normalization. Fig.10 presents the normalized intensity as a function of angle for all scintillators, energies, and ion species used in this study. Despite differences in yield, all scintillators exhibit a qualitatively similar angular trend. This angular dependence is further illustrated in a polar plot representation (Fig.11), which compares the experimental measurements with the proposed cosine-based model and the corresponding optimal fitting parameters. The close agreement between the model and the experimental data demonstrates the suitability of the proposed empirical law for describing the directional dependence of scintillation emission. The ability to quantitatively characterize and model the angular dependence of light output has significant implications for applications in nuclear fusion diagnostics, particularly for estimating the absolute flux of fast ion losses in a nuclear fusion reactor.

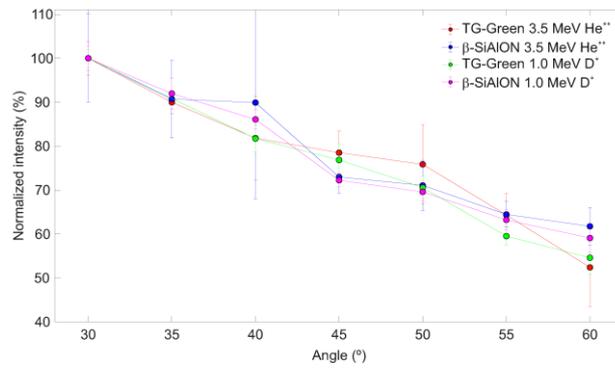

Figure 10: Normalized efficiency as a function of detection angle, showing a consistent decrease across scintillator materials, ion species, and beam energies.

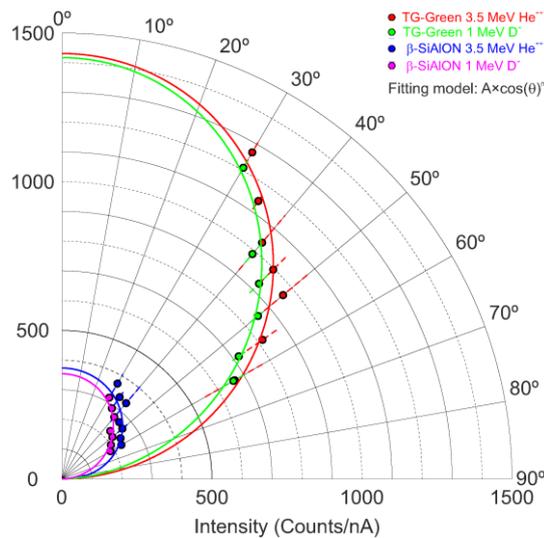

Figure 11: Polar plot of the experimental emission efficiency for TG-Green and β-SiAlON scintillators under different irradiation conditions. Solid lines represent the best fit using a cosine power-law model to the experimental data.

## 5. Conclusions:

This study presents a thorough investigation of the angular emission characteristics of commercial scintillator materials (TG-Green and β-SiAlON) under ion irradiation conditions relevant for nuclear fusion applications. Using 3.5 MeV He$^{++}$ and 1 MeV D$^+$ beams, we quantified the relative ion yield of these materials as a function of observation angle, leveraging a novel experimental setup specifically designed to allow precise optical alignment, angular scanning, and signal normalization. The experimental protocol included rigorous wavelength calibration, evaluation of potential fiber bending losses, and correction for temporal drifts in the light source, ensuring that the observed variations in emission intensity reflect intrinsic material properties rather than extrinsic artifacts. The measurements reveal a pronounced angular anisotropy in scintillation emission for both materials, with light intensity gradually decreasing as the detection angle increases. This angular dependence is consistently observed across different ion species and energies, indicating that the emission behavior is predominantly determined by the scintillator material rather than the incident ion. The observed trend is well captured by an empirical cosine-based model, with exponents close to unity, demonstrating that a first-order cosine law provides a robust and practical approximation for describing the directional emission. TG-Green exhibits a higher light yield compared to β-SiAlON; however, when normalized to beam current, the luminescence response of both materials is comparable, highlighting the reliability of these scintillators for fast-ion diagnostics across multiple irradiation conditions. Under the described irradiation conditions and time intervals, the scintillators exhibited stable light output, with negligible radiation-induced quenching. It should be noted that these tests did not accumulate sufficient fluence to induce significant radiation damage, so this stability reflects performance under the tested conditions rather than long-term radiation hardness. The methodological approach, including the use of cross-correlation techniques to align temporal signals from the ion beam and optical acquisition systems, allowed accurate determination of relative emission yields, addressing limitations present in previous studies that assumed isotropic emission. The characterization of angular emission provides critical insights for improving the accuracy of scintillator-based diagnostics in nuclear fusion devices. In previous analyses, the light collected by the optics was often integrated over the solid angle subtended by the detector, assuming isotropic emission. The present results indicate that this assumption can lead to a systematic underestimation of fast-ion fluxes, as the angular dependence of emission must be explicitly included in such integrals. It is worth noting that in most of the FILDs (AUG, MAST-U, TCV, ITER, …), the optical axis is aligned parallel to the scintillator normal, i.e., observations are made at 0° relative to the surface. In this configuration, the error introduced by angular anisotropy is minimal. However, the current findings highlight that for non-normal lines of sight, which might be considered for space constraints, resolution optimization, or other design requirements, a significant loss of signal can occur. Therefore, the relative orientation between the scintillator and the optics is a key factor to consider for accurate ion flux measurements. Overall, these results establish a robust framework for quantifying the directional response of scintillator materials, bridging a key knowledge gap in ionoluminescence studies and providing practical guidance for the deployment of scintillator-based fast-ion diagnostics in high-performance fusion plasmas.

## Declaration of competing interest

The authors declare that they have no known competing financial interests or personal relationships that could have appeared to influence the work reported in this paper.

## Data availability

The data that has been used is confidential.

## Acknolegments

This research was supported by the French Embassy in Spain, demonstrating their commitment to the "Investigador Consolidado 2025" call. This investigation has also been partially financed by the project ref. ASTRO21/1.1/1 with financing from the European Union-NextGenerationEU, the Ministry of Science, Innovation and Universities, Recovery Plan, Transformation and Resilience, the Department of University, Research and Innovation of the Junta de Andalucía and the University of Sevilla.